\documentclass[11pt]{article}
\usepackage{amsmath}
\usepackage{amssymb}

\usepackage{calc}
\begin{document}
\numberwithin{equation}{section}

\begin{flushright}
EPHOU-14-020 \\
Nov., 2014
\end{flushright}
\vspace*{16mm}

\begin{center}

{\LARGE Off-shell invariant super Yang-Mills with \\ 
gauged central charges for N=D=2 and N=D=4: \\
"Do we need a constraint ?"}\\[8ex]

{\large Noboru Kawamoto
\footnote{Talk given at "Quantum Field Theory and Gravity (QFTG'14)" 
(Tomsk, July28-August3,2014), based on the work in collaboration with 
K.Asaka, K.Nagata and J.Saito\cite{AKNS}.}}\\[4ex]

{\large\itshape
 Graduate School of Science, Hokkaido  University\\
Sapporo, 060-0810 Japan \\[3ex]
}

\end{center}

\hspace{2cm}


\begin{abstract}
We investigate to derive off-shell invariant twisted super Yang-Mills 
for N=2 in 2-dimensions and N=4 in 4-dimensions with a central charge by 
super connection ansatz formalism.  
We find off-shell invariant N=2 algebra with and without an extra 
constraint in 2-dimensions. 
On the other hand in 4-dimensions we find off-shell invariant N=4 
twisted SUSY algebra including one central charge always with a 
constraint. 

\end{abstract}


\newpage

\section{Introduction}

It has been a long-standing question: "Can we construct off-shell 
invariant N=D=4 super Yang-Mills formulation ?" The answer was 
claimed to be negative especially for the case of R-symmetry 
SU(4) case where only on-shell invariance was realized\cite{only on-shell}. 
It was, however, claimed later that off-shell invariance was realized 
for R-symmetry USp(4) case with a central charge\cite{central charge,SSW}. 
However there appeared a constraint in this 
formulation\cite{solving constraint}. 

There were intensive investigations on N=2 and N=4 SUSY algebra with 
central charge with a hope that N=D=4 super Yang-Mills 
can be formulated by superspace formalism\cite{VT}. 
There were also trials by harmonic superspace approach on this 
question\cite{harmonic,Buchbinder}. 

In the analyses of extended SUSY algebra it has been recognized that 
the quantization of gauge theory leads to a twisted version of 
SUSY algebra. It was especially shown that N=2 super Yang-Mills 
in 4-dimensions can be derived by quantizing topological Yang-Mills 
with instanton gauge fixing\cite{topological twist}. 
It has been intensively 
investigated to find a procedure of extending this formulation 
into N=4 super Yang-Mills formulation\cite{Labastida}.

In dealing with N=D=4 supersymmetry algebra we proposed twisted 
superspace formulation by Dirac-Kaehler twisting 
procedure\cite{Dirac-Kaehler, KKM,Saito}. 
This Dirac-Kaehler twist is equivalent to Marcus twist of N=4 in 
4-dimensions\cite{B model} among other twisting 
procedures\cite{Yamron,Vafa-Witten}. 
The Dirac-Kaehler twisting procedure, however, has nice generalization 
to other dimensions. Especially for 2-dimensional N=2 super Yang-Mills 
with central charge we found super connection ansatz where off-shell 
invariant super Yang-Mills can be formulated with and without a 
constraint\cite{D=N=2}. 

It has already been formulated as a super connection ansatz 
where N=D=4 twisted super Yang-Mills with a central charge can be 
formulated at the off-shell level with a constraint\cite{Saito}. 
It is an interesting question to ask if we can 
formulate N=4 super Yang-Mills in 4-dimensions without constraint 
by imposing the similar ansatz as 2-dimensional N=2 case leading 
to a super Yang-Mills formulation without constraint. 

Throughout of this paper we use Euclidean formulation of SUSY 
algebra since we have in mind the application of the formulation 
into lattice SUSY\cite{link-latSUSY, Hopf-Alg-link-app, super-doub}.

\section{Dirac-Kaehler twisted supersymmetry}

We first show how N=D=2 twisted SUSY algebra naturally appears from 
a quantization of gauge theory. 
Let's first consider a very simple 2-dimensional abelian BF theory:
\begin{equation}
S=\int d^2x  \phi\epsilon^{\mu\nu}\partial_\mu\omega_\nu, 
\end{equation}
which has the following gauge symmetry:
\begin{equation}
\phi =0,\ \ 
\delta \omega_\mu =\partial_\mu v.
\end{equation}
After the Lorentz gauge fixing: $\partial^\mu\omega_\mu=0$, 
a quantized action leads:
\begin{equation} 
S=\int d^2x 
 [\epsilon^{\mu\nu}\phi\partial_\mu\omega_\nu
 +b\partial^\mu\omega_\mu
 -i\bar{c}\partial^\mu \partial_\mu c ],
\end{equation}
which has BRST invariance with nilpotent BRST charge $s^2=0$. 
It is interesting to recognize that we can find family of 
BRST charges $s_\mu$ and $\tilde{s}$ which has the following 
fermionic symmetry at the on-shell level:
\begin{center}
\begin{tabular}{c|c|c|c}
\hline
$\phi^A$ & $s\phi^A$ & $s_\mu \phi^A$ & $\widetilde{s}\phi^A$ \\
\hline
$\phi$
 & $0$
 & $-\epsilon_{\mu\nu}\partial^\nu\bar{c}$ 
 & $0$ \\
$\omega_\nu$
 & $\partial_\nu c$
 & $0$
 & $-\epsilon_{\nu\rho}\partial^\rho c$\\
$c$
 & $0$
 & $-i\omega_\mu$
 & $0$\\
$\bar{c}$
 & $-ib$
 & $0$
 & $-i\phi$\\
$b$
 & $0$
 & $\partial_\mu\bar{c}$
 & $0$\\
\hline
\vspace{0.5cm}
\end{tabular}\\
On-shell N=D=2 twisted supersymmetry.
\end{center}
In fact these family of femionic charges satisfy the following 
twisted N=D=2 supersymmetry algebra:
\begin{equation}
\{s,s_\mu\}=-i\partial_\mu,~~\{\tilde{s},s_\mu\}=
i\epsilon_{\mu\nu}\partial^\nu, ~~~~~~
s^2=\{s,\tilde{s}\}=\tilde{s}^2=\{s_\mu,s_\nu\}=0.
\end{equation}

What is surprising here is that we can find off-shell invariant 
N=D=2 supersymmetric action by introducing auxiliary fields 
$\lambda$ and $\rho$: 
\begin{align*}
S&=\int d^2x 
 [\epsilon^{\mu\nu}\phi\partial_\mu\omega_\nu
 +b\partial^\mu\omega_\mu
 -i\bar{c}\partial^\mu \partial_\mu c 
-i\lambda\rho] \\
&=\int d^2x 
s\tilde{s}\frac{1}{2}\epsilon^{\mu\nu}s_\mu s_\nu (-i\bar{c}c), 
\end{align*}
which has the s-exact form of the action with respect to the super 
charges. It has the following off-shell invariant twisted 
N=D=2 supersymmetry: 
\begin{center}
\begin{tabular}{c|c|c|c}
\hline
$\phi^A$ & $s\phi^A$ & $s_\mu \phi^A$ & $\widetilde{s}\phi^A$ \\
\hline
$\phi$
 & $i\rho$
 & $-\epsilon_{\mu\nu}\partial^\nu\bar{c}$ 
 & $0$ \\
$\omega_\nu$
 & $\partial_\nu c$
 & $-i\epsilon_{\mu\nu}\lambda$
 & $-\epsilon_{\nu\rho}\partial^\rho c$\\
$c$
 & $0$
 & $-i\omega_\mu$
 & $0$\\
$\bar{c}$
 & $-ib$
 & $0$
 & $-i\phi$\\
$b$
 & $0$
 & $\partial_\mu\bar{c}$
 & $-i\rho$\\
$\lambda$
 & $\epsilon^{\mu\nu}\partial_\mu\omega_\nu$
 & $0$
 & $-\partial_\mu\omega^\mu$\\
   $\rho$
 & $0$
 & $-\partial_\mu\phi-\epsilon_{\mu\nu}\partial^\nu b$
 & $0$\\
\hline
\end{tabular}\\
\vspace{0.5cm}
Off-shell invariant N=D=2 supersymmetry.
\end{center}

We call this fermionic symmetry algebra as twisted supersymmetry 
since the fermionic charges $s_A$ are related with super charges 
of N=D=2 super symmetry algebra in the following way:
\begin{equation}
\{Q_{\alpha i},Q_{\beta j} \}
 =2\delta_{ij} {\gamma^\mu}_{\alpha\beta}P_\mu, ~~
Q_{\alpha i}= \left(\mathbf{1} s + \gamma^\mu s_\mu  -i \gamma^5
\tilde{s}\right)_{\alpha i}, 
\end{equation}
where $\gamma^\mu$ and $\gamma^5$ are properly chosen 
$\gamma$-matrices in 
2-dimensions\cite{D=N=2}. We call this 
supersymmetry algebra as Dirac-Kaehler twisted supersymmetry 
algebra\cite{Dirac-Kaehler,KKM,Saito}. This Dirac-Kaehler twisting 
procedure can be extended 
into 4-dimensions for N=D=4 supersymmetry algebra\cite{KKM,Saito}.

We can extend the N=D=2 twisted supersymmetry algebra to include 
super charges\cite{D=N=2}:
\begin{align} 
\{Q_{\alpha i},Q_{\beta j}\} =2\delta_{ij}\gamma^{\mu}_{\alpha\beta}P_{\mu}+
2\delta_{\alpha\beta}\delta_{ij}U_{
0}+2\gamma^{5}_{\alpha\beta}\gamma^{5}_{ij}V_{5}, \\
[Q_{\alpha i},R] =iS_{ij}Q_{\alpha j},~~~ 
[U_{0},\hbox{any}] =[V_{5},\hbox{any}]=0,
\end{align}
where $R$ is the R-symmetry generator and $U_0$ and $V_5$ are considered 
as central charges. Then twisted supersymmetry algebra with central 
charges is given by 
\begin{equation}
\begin{split}
&\{s,s_{\mu}\}=P_{\mu}\,,
 \ \{\tilde{s},s_{\mu}\}=-\epsilon_{\mu\nu}P_{\nu}\,,
 \ \{s,\tilde{s}\}=0,\\
&s^{2}=\tilde{s}^{2}=\frac{1}{2}(U_{0}-V_{5})\,,
 \ \{s_{\mu},s_{\nu}\}=\delta_{\mu\nu}(U_{0}+V_{5}).
\label{ts-alg-wc}
\end{split}
\end{equation}

\section{Twisted superspace and super connection}

We introduce a twisted super field which is expanded by super 
coordinates $\theta_A$ corresponding to super charge $s_A$: 
\begin{equation}
\Phi(x_{\mu},\theta_{A},z)=\phi(x_{\mu},z)+\theta_{A}\phi_{A}(x_{\mu},z)+
\frac{1}{2}\theta_{A}\theta_{B}\phi_{AB}(x_{\mu},z)+\cdots, 
\end{equation}
where $x_\mu$ is the space time coordinate and $z$ is a coordinate 
(parameter) corresponding to a central charge. 
Twisted supersymmetry transformation is generated by 
super chages $Q_A$ and we introduce super derivative $\mathcal{D}_{J}$ as:
\begin{equation} 
\delta_{\xi}\Phi=\xi_{A}\mathcal{Q}_{A}\Phi,  ~~
\{\mathcal{Q}_{I},\mathcal{D}_{J}\}=0, 
\end{equation}
where $\xi_{A}$ is super parameters. 
In order to consider gauge theory in this twisted superspace we introduce 
super covariant derivative 
\begin{equation}
\nabla_{I}=\mathcal{D}_{I}-i\Gamma_{I}~~ (I=A),
\label{super-connection}
\end{equation}
where $\Gamma_{I}$ can be identified as super connection. 
We introduce a notation to express the lowest order term 
with respect to the super coordinates; i.e.  
$\Phi|_{\theta_A=0}=\Phi|\equiv\phi$. 
If we generalize the notation of (\ref{super-connection}) for 
the gauge covariant derivative with $I=\underline{\mu},z$ then 
$\nabla_{\underline{\mu}}| \equiv D_{\mu}=\partial_{\mu}-iA_{\mu}$. 

Let us introduce a table notation of (anti-)commuting relations. 
For example the twisted supersymmetry algebra with central charge 
in (\ref{ts-alg-wc}) can be read as \\
\begin{center}
\begin{tabular}{c|c|c|c}
&$s$&$s_\mu$&$\tilde{s}$ \\ \hline
$s$&$\frac{1}{2}(U_{0}-V_{5})$&$P_\mu$&0  \\
$s_\nu$& &$\delta_{\nu\mu}(U_{0}+V_{5})$& $-\epsilon_{\nu\rho} P^\rho$ \\
$ \tilde{s}$&&&$\frac{1}{2}(U_{0}-V_{5})$ \\
\end{tabular}
\end{center}
where a low and a column crossing location of term represents the 
value of the corresponding anti-commutation relation of super charges. 

\section{Super connection ansatz and SUSY transformation}

Once SUSY algebra is given, it is straightforward to examine the 
full closure of the SUSY algebra by super connection 
formalism\cite{Buchbinder,KKM,Labastida,Milewski}. 
For an ansatz of given SUSY algebra all the possible combinations 
of Jacobi identities give a criteria for a consistency of the 
full algebra. 

Let us consider the following super connection Ansatz (A):\\
\begin{center}
\begin{tabular}{c|c|c|c|c|c}
&$\nabla$&$\tilde{\nabla}$&$\nabla_{\nu}$&$\nabla_{\underline{\nu}}$&$\nabla_{z}$
 \\ \hline
$\nabla$&$-iW+\nabla_{z}$&0&$-i\nabla_{\underline{\nu}}$&$-iF_{\underline{\nu}}$&$iG$ \\
$\tilde{\nabla}$&&$-iW+\nabla_{z}$&$i\epsilon_{\nu\rho}\nabla_{\underline{\rho}}$&$-i\tilde{F_{\underline{\nu}}}$&$i\tilde{G}$ \\
$\nabla_{\mu}$&&&$\pm\delta_{\mu\nu}(iW+\nabla_{z})$&$-iF_{\mu\underline{\nu}}$&$iG_{\mu}$ \\
$\nabla_{\underline{\mu}}$&&&&$-iF_{\underline{\mu}\underline{\nu}}$&$iG_{\underline{\mu}}$\\
 $\nabla_{z}$&&&&&0
\end{tabular}\\
\vspace{0.5cm}
N=D=2 Ansatz (A)
\end{center}
where we included $\nabla_{\underline{\mu}}$ and $\nabla_{z}$ which are 
defined in $\{\nabla_A,\nabla_B\}$ in the table. 
For this ansatz N=D=2 twisted supersymmetry algebra with central 
charge can be read with an identification,
$\nabla_A \rightarrow s_A,~~ W=0,~~\nabla_z \rightarrow Z, ~
-i\nabla_{\underline{\mu}} \rightarrow -i\partial_\mu = P_\mu$: 
\[
 \{s,s_{\mu}\}=P_{\mu}\,,
 \ \{\tilde{s},s_{\mu}\}=-\epsilon_{\mu\nu}P_{\nu}\,,
 \ \{s,\tilde{s}\}=0\,,\vspace{-10pt}
\]
\[\hspace*{52pt}
 s^{2}=\tilde{s}^{2}=\frac{1}{2}Z\,,
 \ \{s_{\mu},s_{\nu}\}=\pm\delta_{\mu\nu}Z\,.
\]
We next derive all possible non-trivial relations by using graded 
Jacobi identities until we don't get any new relation. 
In other words if we get inconsistent relations from the Jacobi 
identities we consider that the starting super connection ansatz 
is not taken correctly. 

For example the following graded Jacobi identity is satisfied for 
boson W, fermion $\psi$ and fermion $\chi$: 
\begin{equation}
[W,\{\psi,\chi\}] + \{\psi,[\chi,W]\} -\{\chi,[W,\psi]\} = 0. 
\end{equation}
As a concrete example of deriving a non-trivial relation is 
\begin{equation}
[\nabla,\{\nabla, \tilde{\nabla}\}] + [\nabla,\{\tilde{\nabla},\nabla \}] 
+ [\tilde{\nabla},\{\nabla,\nabla \}] =0, 
\end{equation}
where $\{\nabla, \tilde{\nabla}\}=0$. We then obtain the following 
relation:
\begin{equation}
[\tilde{\nabla}, -iW +\nabla_z] = -i\tilde{\nabla} W +i\tilde{G}=0 
\rightarrow \tilde{\nabla} W= \tilde{G}.
\end{equation}
Similarly we obtain the following relations:
\begin{center}
\begin{equation}
\nabla_{\mu}\nabla
 W=\epsilon_{\mu\nu}\nabla_{\nu}\tilde{\nabla}W\,,\nonumber
\end{equation}
\begin{equation}
F_{\underline{\mu}}=-i\nabla_{\mu}W\,,\hspace*{10pt}
\tilde{F}_{\underline{\mu}}=-i\epsilon_{\mu\nu}\nabla_{\nu}W\,,\nonumber
\end{equation}
\begin{equation}
F_{\mu\underline{\nu}}=\pm i\delta_{\mu\nu}\nabla W \mp i\epsilon_{\mu\nu}\tilde{\nabla}W\,,\hspace*{10pt}
F_{\underline{\mu}\underline{\nu}}=\pm\epsilon_{\mu\nu}\tilde{\nabla}\nabla
W+\frac{1}{2}\epsilon_{\mu\nu}\epsilon_{\rho\sigma}\nabla_{\rho}\nabla_{\sigma}W\,,\nonumber
\end{equation}
\begin{equation}
G=\nabla W\,,\hspace*{10pt}
\tilde{G}=\tilde{\nabla}W\,,\hspace*{10pt}
G_{\mu}=-\nabla_{\mu}W\,,\hspace*{10pt}
G_{\underline{\mu}}=2i\nabla_{\mu}\nabla W-\nabla_{\underline{\mu}}W.\nonumber
\end{equation}
\end{center}
We identify component fields of super multiplets as:
\begin{equation}
W|=\phi\,,\hspace*{10pt}
\nabla W|=\rho\,,\hspace*{10pt}
\tilde{\nabla}W|=\tilde{\rho}\,,\hspace*{10pt}
\nabla_{\mu}W|=\lambda_{\mu}\,,\hspace*{10pt}
\nabla_{z}W|=D\,,\hspace*{10pt}
G_{\underline{\mu}}|=g_{\mu} \nonumber
\end{equation}
SUSY transformations of these component fields can be obtained by 
taking (anti-)commutator with the field and 
identifying the original Jacobi identities. For example the following 
SUSY transformation can be identified as 
\begin{equation}
s\phi=s(W|)\equiv\mathcal{Q} W|=\mathcal{D}W|=\mathcal{D}W|-i[\Gamma,W]| 
=\nabla W| =\rho,
\end{equation}
where Wess-Zumino gauge is chosen here: $\Gamma=0$. 
In a similar way we can derive all the N=D=2 SUSY transformations for 
these component fields: \\
\begin{tabular}{c|c|c|c|c}
&$s$&$s_{\mu}$&$\tilde{s}$&$Z$ \\ \hline
 $\phi$&$\rho$&$\lambda_{\mu}$&$\tilde{\rho}$&$D$ \\[2pt]
 $A_{\nu}$&$-i\lambda_{\nu}$&$\pm i\delta_{\mu\nu}\rho\mp i\epsilon_{\mu\nu}\tilde{\rho}$&$-i\epsilon_{\nu\rho}\lambda_{\rho}$&$g_{\nu}$\\[2pt]
$\lambda_{\nu}$&$\frac{i}{2}(g_{\nu}-D_{\nu}\phi)$&$\pm\frac{1}{2}\delta_{\mu\nu}D+\frac{1}{2}F_{\mu\nu}$&$-\frac{i}{2}\partial_{\nu\rho}(g_{\rho}-D_{\rho}\phi)$&$-iD_{\nu}\rho+i\epsilon_{\nu\rho}D_{\rho}\tilde{\rho}-i[\phi,\lambda_{\nu}]$\\[2pt]
 $\rho$&$\frac{1}{2}D$&$-\frac{i}{2}(g_{\mu}+D_{\mu}\phi)$&$\mp\frac{1}{4}\epsilon_{\mu\nu}F_{\mu\nu}$&$\mp
 iD_{\mu}\lambda_{\mu}+i[\phi,\rho]$\\[2pt]
 $\tilde{\rho}$&$\pm\frac{1}{4}\epsilon_{\mu\nu}F_{\mu\nu}$&$\frac{i}{2}\epsilon_{\mu\nu}(g_{\nu}+D_{\nu}\phi)$&$\frac{1}{2}D$&$\mp
 i\epsilon_{\mu\nu}D_{\mu}\lambda_{\nu}+i[\phi,\tilde{\rho}]$\\[2pt]
 $D$&$\mp iD_{\mu}\lambda_{\mu}$&$i\epsilon_{\mu\nu}D_{\nu}\tilde{\rho}-iD_{\mu}\rho$&$\mp
 i\epsilon_{\mu\nu}D_{\mu}\lambda_{\nu}$&$\pm D_{\mu}g_{\mu}\mp
 D_{\mu}D_{\mu}\phi$\\
&&&&$\pm 2i\{\lambda_{\mu},\lambda_{\mu}\}+i[\phi,D]$\\[2pt]
 $g_{\nu}$&$\epsilon_{\nu\rho}D_{\rho}\tilde{\rho}-[\phi,\lambda_{\nu}]$&$\epsilon_{\mu\sigma}\epsilon_{\nu\rho}D_{\rho}\lambda_{\sigma}$&$-\epsilon_{\nu\rho}(D_{\rho}\rho+[\phi,\lambda_{\rho}])$&$\pm
 D_{\rho}F_{\nu\rho}-2\epsilon_{\nu\rho}\{\lambda_{\rho},\tilde{\rho}\}$\\[2pt]
 &&$\mp\delta_{\mu\nu}[\phi,\rho]\pm\epsilon_{\mu\nu}[\phi,\tilde{\rho}]$&&$-2\{\lambda_{\nu},\rho\}+i[\phi,D_{\nu}\phi]$
\end{tabular}
\vspace{0.5cm}
\begin{center}
N=D=2 SUSY transformation for Ansatz (A).
\end{center}

We can find off-shell invariant action under this SUSY transformation as:\\
\begin{align}
S=\displaystyle\int
 d^{2}x\text{Tr}\Bigl(&\pm\dfrac{1}{2}(D_{\mu}\phi)^{2}
-\frac{1}{4}F^{2}_{\mu\nu}-\frac{1}{2}D^{2}\pm\frac{1}{2}g^{2}_{\mu}
\mp2i\lambda_{\mu}(D_{\mu}\rho-\epsilon_{\mu\nu}D_{\nu}\tilde{\rho})
\nonumber\\
&-i\phi\{\rho,\rho\}-i\phi\{\tilde{\rho},\tilde{\rho}\}\pm 
i\phi\{\lambda_{\mu},\lambda_{\mu}\}\Bigr). \nonumber
\end{align}
In order to confirm off-shell closure of the algebra we need the 
following non-trivial constraint:
\begin{equation}
iD_{\mu}g_{\mu}\mp[\phi,D]-\{\lambda_{\mu},\lambda_{\mu}\}
\mp\{\rho,\rho\}\mp\{\tilde{\rho},\tilde{\rho}\}=0,
\label{constraint-2d}
\end{equation}
which cannot be obtained as one of Jacobi identities. 
For Abelian case the constraint becomes simple as: 
$\partial_{\mu}g_{\mu}=0$, and can be solved as 
$g_\mu=\epsilon_{\mu\nu}\partial_\nu B$. We can then obtain 
off-shell SUSY invariant action without constraint:
\begin{align}
S=\displaystyle\int
 d^{2}x\text{Tr}\Bigl(&\pm\dfrac{1}{2}(\partial_{\mu}\phi)^{2}
-\frac{1}{4}F^{2}_{\mu\nu}-\frac{1}{2}D^{2}\pm\frac{1}{2}(\partial_{\mu}B)^{2}\mp2i\lambda_{\mu}(\partial_{\mu}\rho-\epsilon_{\mu\nu}\partial_{\nu}\tilde{\rho})\nonumber\\
&+e(\frac{1}{2}\phi\epsilon_{\mu\nu}F_{\mu\nu}+2\rho\tilde{\rho}+BD+\epsilon_{\mu\nu}\lambda_{\mu}\lambda_{\nu})\Bigr). 
\nonumber
\end{align}
The constraint (\ref{constraint-2d}) cannot be solved for the 
non-Abelian case in a local way. This example is similar to the 
N=D=4 super Yang-Mills with R-symmetry USp(4) case where one constraint 
appears as a extra condition.

\section{Off-shell N=D=2 SUSY invariant action without constraint}

In order to find other ansatz which doesn't generate a constrained 
equation like (\ref{constraint-2d}), we impose another ansatz as in 
the following:\\
\begin{center}
\begin{tabular}{c|c|c|c|c|c}
&$\nabla$&$\tilde{\nabla}$&$\nabla_{\nu}$&$\nabla_{\underline{\nu}}$&$\nabla_{z}$
 \\ \hline
$\nabla$&$0$&0&$-i(\nabla_{\underline{\nu}}+F_{\nu})$&$-iF_{\underline{\nu}}$&$iG$ \\
$\tilde{\nabla}$&&$0$&$i\epsilon_{\nu\rho}(\nabla_{\underline{\rho}}-F_{\rho})$&$-i\tilde{F_{\underline{\nu}}}$&$i\tilde{G}$ \\
$\nabla_{\mu}$&&&$\delta_{\mu\nu}\nabla_{z}$&$-iF_{\mu\underline{\nu}}$&$iG_{\mu}$ \\
$\nabla_{\underline{\mu}}$&&&&$-iF_{\underline{\mu}\underline{\nu}}$&$iG_{\underline{\mu}}$\\
 $\nabla_{z}$&&&&&0
\end{tabular}\\
\vspace{0.5cm}
N=D=2 Ansatz (B)
\end{center}
From this ansatz we obtain the following relations by using graded 
Jacobi identities:
\begin{center}
\begin{align}
&\nabla F_{\mu}=\epsilon_{\mu\nu}\tilde{\nabla}F_{\nu}\,,\hspace*{10pt}
\nabla_{\mu}F_{\nu}+\nabla_{\nu}F_{\mu}=\delta_{\mu\nu}\nabla_{\rho}F_{\rho}\,,\hspace*{10pt}
G_{\mu}=0\,,\nonumber\\
&\hspace*{70pt}F_{\underline{\mu}}=-i\nabla F_{\mu}\,,\hspace*{10pt}
\tilde{F}_{\underline{\mu}}=i\tilde{\nabla}F_{\mu}\,,\nonumber\\
&\hspace*{20pt}F_{\mu\underline{\nu}}=-\frac{i}{2}\delta_{\mu\nu}(\nabla_{\rho}F_{\rho}-G)+
\frac{i}{2}\epsilon_{\mu\nu}(\epsilon_{\rho\sigma}\nabla_{\rho}F_{\sigma}-\tilde{G})\,,\nonumber\\
&\hspace*{20pt}F_{\underline{\mu}\underline{\nu}}=\nabla_{\mu}\nabla
 F_{\nu}-\nabla_{\nu}\nabla F_{\mu}+i[F_{\mu},F_{\nu}]+\frac{1}{2}\epsilon_{\mu\nu}\nabla\tilde{G}\,,\nonumber\\
&\hspace*{68pt}\nabla G=\tilde{\nabla}\tilde{G}=\nabla\tilde{G}+\tilde{\nabla}G=0\,,\nonumber\\
&\nabla_{z}F_{\mu}=\frac{1}{2}(\nabla_{\mu}G-\epsilon_{\mu\nu}\nabla_{\nu}\tilde{G})\,,\hspace*{10pt}
G_{\underline{\mu}}=\frac{i}{2}(\nabla_{\mu}G+\epsilon_{\mu\nu}\nabla_{\nu}\tilde{G})\,\label{const-relations-(2d-B)}.
\end{align}
\end{center}
In this ansatz we make the following identification of component fields 
of N=D=2 super multiplet:
\begin{equation}
F_{\mu}|=\phi_{\mu}\,,\hspace*{10pt}
\nabla F_{\mu}|=\lambda_{\mu}\,,\hspace*{10pt}
\nabla_{\mu}F_{\nu}|=\frac{1}{2}(\delta_{\mu\nu}\rho+
\epsilon_{\mu\nu}\tilde{\rho})\,,\hspace*{10pt}
\nabla_{\mu}\nabla F_{\mu}|=D.
\end{equation}
SUSY transformation of these component fields can be obtained as in 
the previous example: 
\begin{center}
\begin{tabular}{c|c|c|c|c}
 &$s$&$s_{\mu}$&$\tilde{s}$&$Z$ \\ \hline
 $\phi_{\nu}$&$\lambda_{\nu}$&$\frac{1}{2}(\delta_{\mu\nu}\rho+\epsilon_{\mu\nu}\tilde{\rho})$&$-\epsilon_{\nu\rho}\lambda_{\rho}$&$\frac{1}{2}(\nabla_{\nu}G|-\epsilon_{\nu\rho}\nabla_{\rho}\tilde{G}|)$  \\[2pt]
 $A_{\nu}$&$-i\lambda_{\nu}$&$-\frac{i}{2}\delta_{\mu\nu}\rho+\frac{i}{2}\epsilon_{\mu\nu}\tilde{\rho}$&$-i\epsilon_{\nu\rho}\lambda_{\rho}$&$\frac{i}{2}(\nabla_{\nu}G|+\epsilon_{\nu\rho}\nabla_{\rho}\tilde{G}|)$ \\
&&$+\frac{i}{2}\delta_{\mu\nu}G|-\frac{i}{2}\epsilon_{\mu\nu}\tilde{G}|$&&\\[2pt]
$\lambda_{\nu}$&$0$&$A_{\mu\nu}$&$0$&$-\frac{i}{2}(D^{-}_{\nu}G|-\epsilon_{\nu\rho}D^{+}_{\rho}\tilde{G}|)$ \\[2pt]
$\rho$&$\frac{i}{2}[D^{+}_{\rho},D^{-}_{\rho}]-D$&$\frac{1}{2}(\nabla_{\mu}G|-\epsilon_{\mu\nu}\nabla_{\nu}\tilde{G}|)$&$\frac{i}{2}\epsilon_{\rho\sigma}[D^{-}_{\rho},D^{-}_{\sigma}]$&$\frac{1}{2}(\nabla_{z}G|-\epsilon_{\rho\sigma}\nabla_{\rho}\nabla_{\sigma}\tilde{G}|)$ \\[2pt]
$\tilde{\rho}$&$-\frac{i}{2}\epsilon_{\rho\sigma}[D^{+}_{\rho},D^{+}_{\sigma}]$&$\frac{1}{2}(\nabla_{\mu}\tilde{G}|+\epsilon_{\mu\nu}\nabla_{\nu}G|)$&$-\frac{i}{2}[D^{+}_{\rho},D^{-}_{\rho}]-D$&$\frac{1}{2}(\nabla_{z}\tilde{G}|+\epsilon_{\rho\sigma}\nabla_{\rho}\nabla_{\sigma}G|)$  \\[2pt]
$D$&$-iD^{+}_{\rho}\lambda_{\rho}$&$\frac{i}{2}(D^{+}_{\mu}\rho-\epsilon_{\mu\nu}D^{-}_{\nu}\tilde{\rho})$&$-i\epsilon_{\rho\sigma}D^{-}_{\rho}\lambda_{\sigma}$&$-\frac{i}{2}(D^{-}_{\rho}\nabla_{\rho}G|+\epsilon_{\rho\sigma}D^{+}_{\rho}\nabla_{\sigma}\tilde{G}|)$ \\
&&$-\frac{i}{2}(D_{\mu}G|-\epsilon_{\mu\nu}D_{\nu}\tilde{G}|)$&&$i\{\rho,G|\}+i\{\tilde{\rho},\tilde{G}|\}$ \\
&&&&$-\frac{i}{2}(\{G|,G|\}+\{\tilde{G}|,\tilde{G}|\})$
\end{tabular}\\
\vspace{0.5cm}
N=D=2 SUSY transformation for Ansatz (B).
\end{center}

In this ansatz the super charges have the following nilpotent nature:
\begin{equation}
s^{2}=\tilde{s}^{2}=s_{\pm}^{2}=0, ~~(s_{\pm}\equiv s_{1}\pm is_{2}).
\end{equation}
In order to derive off-shell SUSY invariant action it is convenient 
to find a s-exact form of an action. To find this type of action 
we recognized that among the relations in 
(\ref{const-relations-(2d-B)}) the relations on $G$ and $\tilde{G}$ 
are crucial to be solved. We actually found a solution of 
$\nabla\tilde{G}=\tilde{\nabla}G=0$ as:
\begin{equation} 
G=a\epsilon_{\mu\nu}\nabla^{-}_{\underline{\mu}}\nabla F_{\nu}\,,\hspace*{10pt}
\tilde{G}=-a\nabla^{+}_{\underline{\mu}}\nabla F_{\mu},
\end{equation}
where $a$ is a constant. 
We then found a off-shell SUSY invariant action:\\
\begin{equation}
S=\displaystyle\int
 d^{2}x\hbox{Tr}\{\frac{1}{2}(D_{\mu}\phi_{\nu})^{2}+
\frac{1}{4}F^{2}_{\mu\nu}+\frac{1}{2}D^{2}
-i\rho D^{+}_{\mu}\lambda_{\mu}
-i\tilde{\rho}\epsilon_{\mu\nu}D^{-}_{\mu}\lambda_{\nu}
-\frac{1}{4}[\phi_{\mu},\phi_{\nu}]^{2}
-ia^{-1}G|\tilde{G}|\},
\end{equation}
where $G|$ and $\tilde{G}|$ are fermionic fields.
The action satisfies N=2 SUSY invariance at the off-shell level without 
constraint:
\begin{equation}
sS=\tilde{s}S=s_{\pm}S=0.
\end{equation}

\section{N=D=4 super Yang-Mills formulation with central charges}

Most general N=4 supersymmetry algebra in 4-dimensions can be given 
by \\
\begin{align}
  \{Q_{\alpha i},Q_{\beta
 j}\}&=2C^{-1}_{ij}(\gamma^{\mu}C)_{\alpha\beta}P_{\mu}\notag\\
&+2C_{\alpha\beta}(C^{-1}_{ij}U_{0}+(C^{-1}\gamma_{5})_{ij}U_{5})\notag\\
&+2(\gamma_{5}C)_{\alpha\beta}(C^{-1}_{ij}V_{0}+(C^{-1}\gamma_{5})_{ij}V_{5}). 
\label{N=D=4-al}
\end{align}
The N=D=4 super charges $Q_{\alpha i}$ can be decomposed into 
twisted super charges:
\begin{equation}
Q_{\alpha  i}=\frac{i}{\sqrt{2}}(1s+\gamma^{\mu}s_{\mu}+
\frac{1}{2}\gamma^{\mu\nu}s_{\mu\nu}+
\tilde{\gamma}^{\mu}\tilde{s}_{\mu}+\gamma^{5}\tilde{s})_{\alpha i}.
\end{equation}
In this Dirac-Kaehler twisting mechanism the spinor suffix $\alpha$ 
and the extended SUSY suffix $i$ are rotated by angular momentum 
generators $J_{\mu\nu}$ and R-symmetry generators $R_{\mu\nu}$, 
respectively. In this way the fermionic super charges having spinor 
suffix change into the twisted super charges having scalar, vector, 
tensor.. suffix which are now rotated by a new rotation generators 
$J^{\prime}_{\mu\nu}$. They have the following 
relation\cite{Dirac-Kaehler,KKM,Saito}:
\begin{equation}
J^{\prime}_{\mu\nu}\equiv J_{\mu\nu}+R_{\mu\nu}.
\end{equation}

The N=D=4 Dirac-Kaehler twisted SUSY algebra corresponding to 
(\ref{N=D=4-al}) is given by
\[
 \{s,s_{\mu}\}=\{\tilde{s},\tilde{s}_{\mu}\}=P_{\mu}\,,\vspace{-2pt}
\]
\[
 \{s_{\mu},s_{\rho\sigma}\}=-\delta_{\mu\nu\rho\sigma}P_{\nu}\,,
 \ \{\tilde{s}_{\mu},s_{\rho\sigma}\}=\epsilon_{\mu\nu\rho\sigma}P_{\nu}\,,\vspace{-2pt}
\]
\[
 \{s,\tilde{s}_{\mu}\}=\{\tilde{s},s_{\mu}\}=\{s,s_{\mu\nu}\}=\{\tilde{s},s_{\mu\nu}\}=0\,,\vspace{-2pt}
\]
\[
 2s^{2}=2\tilde{s}^{2}=U_{0}+V_{5}\,,\vspace{-2pt}
\]
\[
 \{s_{\mu},s_{\nu}\}=\{\tilde{s}_{\mu},\tilde{s}_{\nu}\}=\delta_{\mu\nu}(U_{0}-V_{5})
\,,\vspace{-2pt}
\]
\[
\{s,\tilde{s}\}=U_{5}+V_{0}\,,\ \ 
\{s_{\mu},\tilde{s}_{\nu}\}=\delta_{\mu\nu}(U_{5}-V_{0})\,,\vspace{-2pt}
\]
\[
 \{s_{\mu\nu},s_{\rho\sigma}\}=\delta_{\mu\nu\rho\sigma}(U_{0}+V_{5})-\epsilon_{\mu\nu\rho\sigma}(U_{5}+V_{0})\,,\vspace{-2pt}
\]
where $\delta_{\mu\nu\rho\sigma}\equiv
\delta_{\mu\rho}\delta_{\nu\sigma}-\delta_{\mu\sigma}\delta_{\nu\rho}$ 
and $\epsilon_{\mu\nu\rho\sigma}$ is Euclidean $\epsilon$-tensor. 

We first investigate the case of ansatz where no central charge is 
inserted:\\
\begin{center}
\begin{tabular}{c|c|c|c|c|c|c}
&$\nabla$&$\tilde{\nabla}$&$\nabla_{\rho}$&$\tilde{\nabla}_{\rho}$&
$\nabla_{\rho\sigma}$&$\nabla_{\underline{\rho}}$ \\ \hline
$\nabla$&$0$&$-iW$&$-i(\nabla_{\underline{\rho}}+F_{\rho})$&$0$&$0$&
$-iF_{\underline{\rho}}$ \\
$\tilde{\nabla}$&&$0$&$0$&$-i(\nabla_{\underline{\rho}}-F_{\rho})$&$0$&
$-i\tilde{F_{\underline{\rho}}}$ \\
$\nabla_{\mu}$&&&$0$&$-i\delta_{\mu\rho}F$&
$i\delta_{\mu\nu\rho\sigma}(\nabla_{\underline{\nu}}-F_{\nu})$&
$-iF_{\mu\underline{\rho}}$ \\
$\tilde{\nabla}_{\mu}$&&&&$0$&
$-i\epsilon_{\mu\nu\rho\sigma}(\nabla_{\underline{\nu}}+F_{\nu})$&
$-i\tilde{F}_{\mu\underline{\rho}}$ \\
$\nabla_{\mu\nu}$&&&&&$i\epsilon_{\mu\nu\rho\sigma}W$&
$-iF_{\mu\nu\underline{\rho}}$ \\
$\nabla_{\underline{\mu}}$&&&&&&$-iF_{\underline{\mu}\underline{\rho}}$ 
\end{tabular}\\
\vspace{0.5cm}
N=D=4 Ansatz without central charge (A)
\end{center}
Graded Jacobi identities for this ansatz lead the following relations:\\
\begin{align}
&\hspace*{50pt}\nabla W=\tilde{\nabla}W=\nabla_{\mu\nu}W=0\,,\hspace*{10pt}
\nabla_{\mu}F=\tilde{\nabla}_{\mu}F=0\,,\nonumber\\
&\hspace*{30pt}-\delta_{\mu\nu}\nabla
 F=\tilde{\nabla}_{\mu}F_{\nu}+\tilde{\nabla}_{\nu}F_{\mu}\,,\hspace*{10pt}
\delta_{\mu\nu}\tilde{\nabla}
 F=\nabla_{\mu}F_{\nu}+\nabla_{\nu}F_{\mu}\,,\nonumber\\
&\hspace*{120pt}\nabla_{[\mu}F_{\nu]}=\epsilon_{\mu\nu\rho\sigma}\tilde{\nabla}_{\rho}F_{\sigma}\,,\nonumber\\
&\hspace*{92pt}\nabla_{\mu\nu}F_{\rho}=-\delta_{\mu\nu\rho\sigma}\nabla F_{\sigma}+\epsilon_{\mu\nu\rho\sigma}\tilde{\nabla}F_{\sigma}\,,\nonumber\\
&\hspace*{106pt}F_{\underline{\mu}}=-i\nabla F_{\mu}\,,\hspace*{10pt}
\tilde{F}_{\underline{\mu}}=i\tilde{\nabla}F_{\mu}\,,\nonumber\\
&\hspace*{90pt}\nabla_{\mu}W=-2\tilde{\nabla}F_{\mu}\,,\hspace*{10pt}
\tilde{\nabla}_{\mu}W=2\nabla F_{\mu}\,,\nonumber\\
&F_{\mu\underline{\nu}}=-\frac{i}{2}(\delta_{\mu\nu}\tilde{\nabla}F-\delta_{\mu\nu\rho\sigma}\nabla_{\rho}F_{\sigma})\,,\hspace*{10pt}
\tilde{F}_{\mu\underline{\nu}}=-\frac{i}{2}(\delta_{\mu\nu}\nabla
 F+\delta_{\mu\nu\rho\sigma}\tilde{\nabla}{\rho}F_{\sigma})\,,\nonumber\\
&\hspace*{90pt}F_{\mu\nu\underline{\rho}}=
-i\delta_{\mu\nu\rho\sigma}\nabla F_{\sigma}-i\epsilon_{\mu\nu\rho\sigma}\tilde{\nabla}F_{\sigma}.\notag
\end{align}
We define component fields of N=4 super multiplets as 
\begin{align}
&\hspace*{40pt}F_{\mu}|=\phi_{\mu}\,,\hspace*{10pt}
W|=A\,,\hspace*{10pt}
F|=B\,,\nonumber\\
&\hspace*{55pt}\nabla F_{\mu}|=\lambda_{\mu}\,,\hspace*{10pt}
\tilde{\nabla}F_{\mu}|=\tilde{\lambda}_{\mu}\,,\nonumber\\
&\nabla_{\mu}F_{\nu}|=\delta_{\mu\nu}\rho+\rho_{\mu\nu}\,,\hspace*{10pt}
\tilde{\nabla}_{\mu}F_{\nu}|=\delta_{\mu\nu}\tilde{\rho}
+\frac{1}{2}\epsilon_{\mu\nu\rho\sigma}\rho_{\rho\sigma}.\notag
\end{align}

We can then obtain on-shell closed SUSY transformation of these fields as:\\
\begin{tabular}{c|c|c|c}
&$s$&$\tilde{s}$&$s_{\mu\nu}$ \\ \hline
 $\phi_{\rho}$&$\lambda_{\rho}$&$\tilde{\lambda}_{\rho}$&$-\delta_{\mu\nu\rho\sigma}\lambda_{\sigma}+\epsilon_{\mu\nu\rho\sigma}\tilde{\lambda}_{\sigma}$ \\[2pt]
 $A_{\rho}$&$-i\lambda_{\rho}$&$i\tilde{\lambda}_{\rho}$&$-i\delta_{\mu\nu\rho\sigma}\lambda_{\sigma}-i\epsilon_{\mu\nu\rho\sigma}\tilde{\lambda}_{\sigma}$\\[2pt]
$A$&$0$&$0$&$0$ \\[2pt]
$B$&$-2\tilde{\rho}$&$2\rho$&$-\epsilon_{\mu\nu\rho\sigma}\rho_{\rho\sigma}$ \\[2pt]
$\lambda_{\rho}$&$0$&$\-\frac{i}{2}D_{\mu}^{-}A$&$-\frac{i}{2}\epsilon_{\mu\nu\rho\sigma}D_{\sigma}^{+}A$\\[2pt]
$\tilde{\lambda}_{\rho}$&$\frac{i}{2}D_{\mu}^{+}A$&$0$&$-\frac{i}{2}\delta_{\mu\nu\rho\sigma}D_{\sigma}^{-}A$ \\[2pt]
 $\rho$&$-\frac{i}{2}(D_{\alpha}\phi_{\alpha}+\frac{1}{2}[A,B])$&$0$&$\frac{i}{2}[D_{\mu}^{-},D_{\nu}^{-}]$\\[2pt]
 $\tilde{\rho}$&$0$&$-\frac{i}{2}(D_{\alpha}\phi_{\alpha}-\frac{1}{2}[A,B])$&$\frac{i}{4}\epsilon_{\mu\nu\rho\sigma}[D_{\mu}^{+},D_{\nu}^{+}]$ \\[2pt]
 $\rho_{\rho\sigma}$&$-\frac{i}{2}[D_{\rho}^{+},D_{\sigma}^{+}]$&$\frac{i}{4}\epsilon_{\mu\nu\rho\sigma}[D_{\mu}^{-},D_{\nu}^{-}]$&$\frac{i}{2}\delta_{\mu\nu\alpha\gamma}\delta_{\rho\sigma\beta\gamma}[D_{\alpha}^{-},D_{\beta}^{+}]-\frac{i}{2}\delta_{\mu\nu\rho\sigma}(D_{\alpha}\phi_{\alpha}+\frac{1}{2}[A,B])$
\end{tabular}\\
\vspace{0.5cm}
\begin{tabular}{c|c|c}
&$s_{\mu}$&$\tilde{s}_{\mu}$\\ \hline
$\phi_{\rho}$&$\delta_{\mu\rho}+\rho_{\mu\rho}$&$\delta_{\mu\rho}\tilde{\rho}+\frac{1}{2}\epsilon_{\mu\rho\alpha\beta}\rho_{\alpha\beta}$\\[2pt]
$A_{\rho}$&$-i\delta_{\mu\rho}+i\rho_{\mu\rho}$&$i\delta_{\mu\rho}\tilde{\rho}-\frac{i}{2}\epsilon_{\mu\rho\alpha\beta}\rho_{\alpha\beta}$\\[2pt]
$A$&$-2\tilde{\lambda}_{\mu}$&$2\lambda_{\mu}$\\[2pt]
$B$&$0$&$0$\\[2pt]
$\lambda_{\rho}$&$\frac{i}{2}[D_{\mu}^{+},D_{\rho}^{-}]+\frac{i}{2}\delta_{\mu\rho}(D_{\alpha}\phi_{\alpha}+\frac{1}{2}[A,B])$&$\frac{i}{4}\epsilon_{\mu\rho\alpha\beta}[D_{\alpha}^{+},D_{\beta}^{+}]$\\[2pt]
$\tilde{\lambda}$&$-\frac{i}{4}\epsilon_{\mu\rho\alpha\beta}[D_{\alpha}^{-},D_{\beta}^{-}]$&$-\frac{i}{2}[D_{\mu}^{-},D_{\rho}^{+}]+\frac{i}{2}\delta_{\mu\rho}(D_{\alpha}\phi_{\alpha}-\frac{1}{2}[A,B])$\\[2pt]
$\rho$&$0$&$-\frac{i}{2}D_{\mu}^{-}B$\\[2pt]
$\tilde{\rho}$&$\frac{i}{2}D_{\mu}^{+}B$&$0$\\[2pt]
$\rho_{\rho\sigma}$&$-\frac{i}{2}\epsilon_{\mu\nu\rho\sigma}D_{\nu}^{-}B$&$\frac{i}{2}\delta_{\mu\nu\rho\sigma}D_{\nu}^{+}B$\\[2pt]
\end{tabular}
\\

We then obtain on-shell invariant N=4 super Yang-Mills action without 
central charge having SU(4) R-symmetry: 
\begin{align}
S=\int
 d^{4}x \hbox{Tr} \bigl\{
&\frac{1}{4}(D_{\mu}\phi_{\nu})^{2}+\frac{1}{8}F_{\mu\nu}^{2}+\frac{1}{8}D_{\mu}^{+}AD_{\mu}^{-}B+\frac{1}{8}D_{\mu}^{-}AD_{\mu}^{+}B-\frac{1}{8}[\phi_{\mu},\phi_{\nu}]^{2}+\frac{1}{16}[A,B]^{2}\nonumber\\
&-i\lambda_{\mu}(D_{\mu}^{+}\rho-D_{\nu}^{-}\rho_{\mu\nu}-[B,\tilde{\lambda}_{\mu}])-i\tilde{\lambda}_{\mu}(D_{\mu}^{-}\tilde{\rho}-\frac{1}{2}\epsilon_{\mu\nu\rho\sigma}D_{\nu}^{+}\rho_{\rho\sigma})\nonumber\\
&+i\tilde{\rho}[A,\rho]-\frac{i}{8}\epsilon_{\mu\nu\rho\sigma}A\{\rho_{\mu\nu},\rho_{\rho\sigma}\}\bigr\} \notag
\end{align}

We now investigate the following super connection ansatz of N=D=4 with a 
central charge: \\
\begin{center}
\begin{tabular}{c|c|c|c|c|c|c|c}
&$\nabla$&$\tilde{\nabla}$&$\nabla_{\rho}$&$\tilde{\nabla}_{\rho}$&$\nabla_{\rho\sigma}$&$\nabla_{\underline{\rho}}$&$\nabla_{z}$ \\ \hline
$\nabla$&$0$&$\nabla_{z}$&$-i(\nabla_{\underline{\rho}}+F_{\rho})$&$0$&$0$&$-iF_{\underline{\rho}}$&$iG$ \\
$\tilde{\nabla}$&&$0$&$0$&$-i(\nabla_{\underline{\rho}}-F_{\rho})$&$0$&$-i\tilde{F_{\underline{\rho}}}$&$i\tilde{G}$ \\
$\nabla_{\mu}$&&&$0$&$\delta_{\mu\rho}(\mp \nabla_{z}-iW)$&$i\delta_{\mu\nu\rho\sigma}(\nabla_{\underline{\nu}}-F_{\nu})$&$-iF_{\mu\underline{\rho}}$&$iG_{\mu}$ \\
$\tilde{\nabla}_{\mu}$&&&&$0$&$-i\epsilon_{\mu\nu\rho\sigma}(\nabla_{\underline{\nu}}+F_{\nu})$&$-i\tilde{F}_{\mu\underline{\rho}}$&$iG_{\mu}$ \\
$\nabla_{\mu\nu}$&&&&&$-\epsilon_{\mu\nu\rho\sigma}\nabla_{z}$&$-iF_{\mu\nu\underline{\rho}}$&$iG_{\mu\nu}$ \\
$\nabla_{\underline{\mu}}$&&&&&&$-iF_{\underline{\mu}\underline{\rho}}$&$iG_{\underline{\mu}}$ \\
$\nabla_{z}$&&&&&&&$0$
\end{tabular}\\
\vspace{0.5cm}
N=D=4 Ansatz with a central charge (B).
\end{center}

The corresponding twisted SUSY algebra with a central charge $Z$ is 
given by 
\begin{align}
\{s,s_{\mu}\}=\{\tilde{s},\tilde{s}_{\mu}\}=P_{\mu}\,,\hspace*{10pt}
\{s_{\mu},s_{\rho\sigma}\}=&-\delta_{\mu\nu\rho\sigma}P_{\nu}\,,\hspace*{10pt}
\{\tilde{s}_{\mu},s_{\rho\sigma}\}=\epsilon_{\mu\nu\rho\sigma}P_{\nu}\,,\nonumber\\
\{s,\tilde{s}_{\mu}\}=\{\tilde{s},s_{\mu}\}=\{s,&s_{\mu\nu}\}=\{\tilde{s},s_{\mu\nu}\}=0\,,\nonumber\\
2s^{2}=2\tilde{s}^{2}=0\,,\hspace*{10pt}
\{s_{\mu},s_{\nu}\}=&\{\tilde{s}_{\mu},\tilde{s}_{\nu}\}=0
\,,\hspace*{10pt}
\{s,\tilde{s}\}=Z\,,\nonumber\\
\{s_{\mu},\tilde{s}_{\nu}\}=\mp\delta_{\mu\nu}Z\,,\hspace*{10pt}
\{&s_{\mu\nu},s_{\rho\sigma}\}=-\epsilon_{\mu\nu\rho\sigma}Z,\notag
\end{align} 
where $ Z=U_5$ for + and $ Z=V_0$ for -. 
Graded Jacobi identities lead:
\begin{center}
\begin{align}
\nabla_{(\mu}F_{\nu)}=\delta_{\mu\nu}\tilde{\nabla}W\,,\hspace*{10pt}
&\tilde{\nabla}_{(\mu}F_{\nu)}=-\delta_{\mu\nu}\nabla W\,,\nonumber\\
\nabla F_{\mu}=\pm\frac{1}{2}\tilde{\nabla}_{\mu}W\,,\hspace*{10pt}
\tilde{\nabla}F_{\mu}=\mp&\frac{1}{2}\nabla_{\mu}W\,,\hspace*{10pt}
\nabla_{\mu\nu}W=-\epsilon_{\mu\nu\rho\sigma}\nabla_{\rho}F_{\sigma}\,,\nonumber\\
\nabla_{[\mu}F_{\nu]}=\epsilon_{\mu\nu\rho\sigma}\tilde{\nabla}_{\rho}F_{\sigma}\,,\hspace*{10pt}
\tilde{\nabla}_{[\mu}F_{\nu]}=\epsilon_{\mu\nu\rho\sigma}\nabla_{\rho}F_{\sigma}&\,,\hspace*{10pt}
\nabla_{\mu\nu}F_{\rho}=-\delta_{\mu\nu\rho\sigma}\nabla F_{\sigma}+\epsilon_{\mu\nu\rho\sigma}\tilde{\nabla}F_{\sigma}\,,\nonumber\\
F_{\underline{\mu}}=-i\nabla F_{\mu}\,,\hspace*{10pt}
\tilde{F}_{\underline{\mu}}=i\tilde{\nabla}F_{\mu}\,,\hspace*{10pt}
&F_{\mu\underline{\nu}}=-i\nabla_{\nu}F_{\mu}\,,\hspace*{10pt}
\tilde{F}_{\mu\underline{\nu}}=i\tilde{\nabla}_{\nu}F_{\mu}\,,\nonumber\\
F_{\mu\nu\underline{\rho}}=-i(\delta_{\mu\nu\rho\sigma}\nabla F_{\sigma}+\epsilon_{\mu\nu\rho\sigma}\tilde{\nabla}F_{\sigma})\,,\hspace*{10pt}
&F_{\underline{\mu}\underline{\nu}}=\nabla_{[\mu}\nabla F_{\nu]}+i[F_{\mu},F_{\nu}]\,,\nonumber\\
G=\tilde{G}=G_{\mu\nu}=0\,,\hspace*{10pt}
G_{\mu}=2\tilde{\nabla}F_{\mu}\,,\hspace*{10pt}
&\tilde{G}_{\mu}=-2\nabla F_{\mu}\,,\hspace*{10pt}
G_{\underline{\mu}}=\frac{i}{2}(\nabla
G_{\mu}+\tilde{\nabla}\tilde{G}_{\mu})\,,\nonumber\\
ZF_{\mu}=\frac{1}{2}(\nabla G_{\mu}
-\tilde{\nabla}\tilde{G}_{\mu})\,,\hspace*{10pt}
&ZW=2i\nabla_{\underline{\mu}}
F_{\mu}+2\nabla\tilde{\nabla}W\notag
\end{align}
\end{center}
We define component fields of N=4 super multiplets:
\begin{align}
F_{\mu}|=\phi_{\mu}\,,\hspace*{10pt}
W|=A\,,\hspace*{10pt}
&\nabla F_{\mu}|=\lambda_{\mu}\,,\hspace*{10pt}
\tilde{\nabla}F_{\mu}|=\tilde{\lambda_{\mu}}\,,\nonumber\\
\nabla_{\mu}F_{\nu}|=\delta_{\mu\nu}\rho+\rho_{\mu\nu}\,,\hspace*{10pt}
\tilde{\nabla_{\mu}}F_{\nu}|=\delta_{\mu\nu}\tilde{\rho}+&\frac{1}{2}\epsilon_{\mu\nu\rho\sigma}\rho_{\rho\sigma}\,,\hspace*{10pt}
\nabla\tilde{\nabla}W|=H\,,\hspace*{10pt}
G_{\underline{\mu}}|=g_{\mu}\,,\hspace*{10pt}
\nabla_{z}F_{\mu}|=H_{\mu}.\notag
\end{align}
N=4 SUSY transformation of these component fields are given by\\
\begin{center}
\begin{tabular}{c|c|c|c|c}
&$s$&$\tilde{s}$&$s_{\mu}$&$\tilde{s}_{\mu}$\\  \hline
 $\phi_{\rho}$&$\lambda_{\rho}$&$\tilde{\lambda}_{\rho}$&$\delta_{\mu\rho}\rho+\rho_{\mu\rho}$&$\delta_{\mu\rho}\tilde{\rho}+\frac{1}{2}\epsilon_{\mu\rho\alpha\beta}\rho_{\alpha\beta}$ \\[2pt]
 $A_{\rho}$&$-i\lambda_{\rho}$&$i\tilde{\lambda}_{\rho}$&$-i\delta_{\mu\rho}\rho+i\rho_{\mu\rho}$&$i\delta_{\mu\rho}\tilde{\rho}-\frac{i}{2}\epsilon_{\mu\rho\alpha\beta}\rho_{\alpha\beta}$ \\[2pt]
$A^{\prime}$&$-\tilde{\rho}$&$\rho$&$\mp\tilde{\lambda}_{\mu}$&$\pm\lambda_[\mu]$ \\[2pt]
$\lambda_{\rho}$&$0$&$g^{+}_{\rho}$&$\frac{i}{2}[D^{+}_{\mu},D^{-}_{\rho}]-\delta_{\mu\rho}H^{\prime}$&$\frac{i}{4}\epsilon_{\mu\rho\alpha\beta}[D^{+}_{\alpha},D^{+}_{\beta}]$ \\[2pt]
$\tilde{\lambda}_{\rho}$&$-g^{-}_{\rho}$&$0$&$-\frac{i}{4}\epsilon_{\mu\rho\alpha\beta}[D^{-}_{\alpha},D^{-}_{\beta}]$&$-\frac{i}{2}[D^{-}_{\mu},D^{+}_{\rho}]+\delta_{\mu\rho}(\frac{i}{2}[D^{-}_{\sigma},D^{+}_{\sigma}]+H^{\prime})$ \\[2pt]
 $\rho$&$H^{\prime}$&$0$&$0$&$\mp g^{+}_{\mu}-iD^{-}_{\mu}A^{\prime}$ \\[2pt]
 $\tilde{\rho}$&$0$&$-\frac{i}{2}[D^{-}_{\mu},D^{+}_{\mu}]-H^{\prime}$&$\pm
 g^{-}_{\mu}+iD^{+}_{\mu}A^{\prime}$&$0$ \\[2pt]
 $\rho_{\rho\sigma}$&$-\frac{i}{2}[D^{+}_{\rho},D^{+}_{\sigma}]$&$\frac{i}{4}\epsilon_{\rho\sigma\alpha\beta}[D^{-}_{\alpha},D^{-}_{\beta}]$&$\epsilon_{\mu\nu\rho\sigma}(\mp g^{+}_{\nu}-iD^{-}_{\nu}A^{\prime})$&$\delta_{\mu\nu\rho\sigma}(\pm g^{-}_{\nu}+iD^{+}_{\nu}A^{\prime})$ \\[2pt]
$H^{\prime}$&$0$&$-iD^{-}_{\mu}\tilde{\lambda}_{\mu}$&$-iD^{+}_{\mu}\rho$&$-\frac{i}{2}\epsilon_{\mu\nu\rho\sigma}D^{+}_{\nu}\rho_{\rho\sigma}$ \\[2pt]
 $g^{+}_{\rho}$&$\mp(\frac{i}{2}\epsilon_{\mu\nu\alpha\beta}D^{+}_{\sigma}\rho_{\alpha\beta}$&$0$&$-i\delta_{\mu\rho}D^{-}_{\nu}\tilde{\lambda}_{\nu}+iD^{-}_{\rho}\tilde{\lambda}_{\mu}$&$-iD^{-}_{\mu}\lambda_{\rho}-i\epsilon_{\mu\rho\alpha\beta}D^{+}_{\alpha}\tilde{\lambda}_{\beta}$\\
&$-iD^{-}_{\rho}\tilde{\rho}+i[A,\lambda_{\rho}])$&&&\\[2pt]
$g^{-}_{\rho}$&$0$&$\pm(iD^{-}_{\sigma}\rho_{\rho\sigma}$&$i\epsilon_{\mu\rho\alpha\beta}D^{-}_{\alpha}\lambda_{\beta}+iD^{+}_{\mu}\tilde{\lambda}_{\rho}$&$i\delta_{\mu\rho}D^{+}_{\nu}\lambda_{\nu}-iD^{+}_{\rho}\lambda_{\mu}$ \\
&&$-iD^{+}_{\rho}\rho+i[A,\tilde{\lambda}_{\rho}])$&&
\end{tabular}\\
\end{center}
\vspace{0.5cm}
\begin{center}
\begin{tabular}{c|c|c}
&$s_{\mu\nu}$&$Z$ \\[2pt] \hline
$\phi_{\rho}$&$-\delta_{\mu\nu\rho\sigma}\lambda_{\sigma}+
\epsilon_{\mu\nu\rho\sigma}\tilde{\lambda}_{\sigma}$&$H_{\rho}$ \\[2pt]
$A_{\rho}$&$-i\delta_{\mu\nu\rho\sigma}\lambda_{\sigma}-
i\epsilon_{\mu\nu\rho\sigma}\tilde{\lambda}_{\sigma}$&$g_{\rho}$ \\[2pt]
$A^{\prime}$&$-\frac{1}{2}\epsilon_{\mu\nu\rho\sigma}\rho_{\rho\sigma}$&
$\frac{i}{2}[D^{-}_{\mu},D^{+}_{\mu}]+H$ \\[2pt]
$\lambda_{\rho}$&$\epsilon_{\mu\nu\rho\sigma}g^{-}_{\sigma}$&
$\mp(-iD^{-}_{\mu}\tilde{\rho}+\frac{i}{2}\epsilon_{\mu\nu\rho\sigma}D^{+}
\rho_{\rho\sigma}+i[A,\lambda_{\mu}])$ \\[2pt]
$\tilde{\lambda}_{\rho}$&$\delta_{\mu\nu\rho\sigma}g^{+}_{\sigma}$&
$\mp(-iD^{+}_{\mu}\rho+iD^{-}_{\nu}\rho_{\mu\nu}+i[A,\tilde{\lambda}_{\mu}])$ 
\\[2pt]
$\rho$&$\frac{i}{2}[D^{-}_{\mu},D^{-}_{\nu}]$&
$-iD^{-}_{\mu}\tilde{\lambda}_{\mu}$ \\[2pt]
$\tilde{\rho}$&$\frac{i}{4}\epsilon_{\mu\nu\rho\sigma}[D^{+}_{\rho},
D^{+}_{\sigma}]$&$-iD^{+}_{\mu}\lambda_{\mu}$ \\[2pt]
$\rho_{\rho\sigma}$&$\frac{i}{2}\delta_{\mu\nu\alpha\gamma}
\delta_{\rho\sigma\beta\gamma}[D^{-}_{\alpha},D^{+}_{\beta}]+
\delta_{\mu\nu\rho\sigma}H^{\prime}$&
$-i\epsilon_{\mu\nu\rho\sigma}D^{-}_{\rho}
\lambda_{\sigma}-iD^{+}_{[\mu}\tilde{\lambda}_{\nu]}$ \\[2pt]
$H^{\prime}$&$iD^{-}_{[\mu}\lambda_{\nu]}$&$iD^{-}_{\mu}g^{-}_{\mu}+
2i\{\lambda_{\mu},\tilde{\lambda}_{\mu}\}$ \\[2pt]
$g^{+}_{\rho}$&$\mp\epsilon_{\mu\nu\rho\sigma}(-iD^{+}_{\sigma}\rho+
iD^{-}_{\alpha}\rho_{\sigma\alpha}+i[A,\lambda_{\sigma}])$&
$\mp i(\frac{i}{2}D^{-}_{\nu}D^{+}_{\nu}D^{-}_{\mu}+
\epsilon_{\mu\nu\rho\sigma}\{\tilde{\lambda}_{\nu},\rho_{\rho\sigma}\}+ $\\
&&$2\{\rho,\lambda\}+D^{-}_{\mu}H^{\prime}+2[A^{\prime},g^{+}_{\mu}])$ \\[2pt]
$g^{-}_{\rho}$&$\mp\delta_{\mu\nu\rho\sigma}(-iD^{-}_{\sigma}\tilde{\rho}+
\frac{i}{2}\epsilon_{\sigma\gamma\alpha\beta}D^{+}_{\gamma}\rho_{\alpha\beta}+
i[A,\lambda_{\sigma}])$&$\mp i(\frac{i}{2}D^{-}_{\nu}D^{+}_{\mu}D^{+}_{\nu}+
2\{\lambda_{\nu},\rho_{\mu\nu}\}+$\\
&&$2\{\tilde{\rho},\tilde{\lambda}_{\mu}\}+
D^{+}_{\mu}H^{\prime}+2[A^{\prime},g^{-}_{\mu}])$ \\[2pt]
\end{tabular}
\\
N=D=4 SUSY transformation with a central charge for Ansatz (B).
\end{center}
For the off-shell closure of the above SUSY algebra we need the 
following constraint:
\begin{equation}
iD_{\mu}g_{\mu}+[\phi_{\mu},H_{\mu}]\mp2\{\rho,\tilde{\rho}\}+2\{\lambda_{\mu},\tilde{\lambda}_{\mu}\}\mp\frac{1}{4}\epsilon_{\mu\nu\rho\sigma}\{\rho_{\mu\nu},\rho_{\rho\sigma}\}\pm\frac{i}{2}D^{+}_{\mu}D^{-}_{\mu}A\pm\frac{1}{2}[A,H]=0.
\label{4d-constraint1}
\end{equation}
Off-shell twisted N=4 SUSY invariant action in this case is given by 
\begin{align}
S=\displaystyle\int
 d^{4}x\text{Tr}\Bigl(&\dfrac{1}{2}D_{\mu}\phi_{\nu}D_{\mu}\phi_{\nu}
+\frac{1}{4}F^{2}_{\mu\nu}\pm\frac{1}{2}(g^{2}_{\mu}+H^{2}_{\mu})+H(iD_{\mu}\phi_{\mu}+\frac{1}{2}H)-\frac{1}{2}(D_{\mu}\phi_{\mu})^{2}\nonumber\\
-&2i\rho
 D^{+}_{\mu}\lambda_{\mu}-2i\tilde{\rho}D^{-}_{\mu}\tilde{\lambda}_{\mu}-2i\rho_{\mu\nu}(D^{-}_{\mu}\lambda_{\nu}+\frac{1}{2}\epsilon_{\mu\nu\rho\sigma}D^{+}_{\rho}\tilde{\lambda}_{\sigma})
-2iA\{\lambda_{\mu},\tilde{\lambda}_{\mu}\}-\frac{1}{4}[\phi_{\mu},\phi_{\nu}]^{2}\Bigr), \notag
\end{align}
where the constraint (\ref{4d-constraint1}) is crucial to prove the 
off-shell invariance of SUSY for this action. 

We now try another N=D=4 twisted SUSY ansatz which has similarity with the 
N=2 Ansatz (B) leading no constraint in 2-dimensions:\\
\begin{center}
\begin{tabular}{c|c|c|c|c|c|c|c}
&$\nabla$&$\tilde{\nabla}$&$\nabla_{\rho}$&$\tilde{\nabla}_{\rho}$&$\nabla_{\rho\sigma}$&$\nabla_{\underline{\rho}}$&$\nabla_{z}$ \\ \hline
$\nabla$&$\nabla_{z}-iW$&$0$&$-i\nabla_{\underline{\rho}}$&$iF_{\rho}$&$0$&$-iF_{\underline{\rho}}$&$iG$ \\
$\tilde{\nabla}$&&$\nabla_{z}-iW$&$-iF_{\rho}$&$-i\nabla_{\underline{\nu}}$&$0$&$-i\tilde{F_{\underline{\rho}}}$&$i\tilde{G}$ \\
$\nabla_{\mu}$&&&$\pm\delta_{\mu\nu}\nabla_{z}$&$0$&$i\delta_{\mu\nu\rho\sigma}\nabla_{\underline{\nu}}+i\epsilon_{\mu\nu\rho\sigma}F_{\nu}$&$-iF_{\mu\underline{\rho}}$&$iG_{\mu}$ \\
$\tilde{\nabla}_{\mu}$&&&&$\pm\delta_{\mu\rho}\nabla_{z}$&$-i\epsilon_{\mu\nu\rho\sigma}\nabla_{\underline{\nu}}+i\delta_{\mu\nu\rho\sigma}F_{\nu}$&$-i\tilde{F}_{\mu\underline{\rho}}$&$iG_{\mu}$ \\
$\nabla_{\mu\nu}$&&&&&$\delta_{\mu\nu\rho\sigma}(\nabla_{z}-iW)$&$-iF_{\mu\nu\underline{\rho}}$&$iG_{\mu\nu}$ \\
$\nabla_{\underline{\mu}}$&&&&&&$-iF_{\underline{\mu}\underline{\rho}}$&$iG_{\underline{\mu}}$ \\
$\nabla_{z}$&&&&&&&$0$
\end{tabular}\\
\vspace{0.5cm}
N=D=4 Ansatz with a central charge (C).
\end{center}
Graded Jacobi identities lead:\\
\begin{align}
\nabla_{(\mu}F_{\nu)}=\delta_{\mu\nu}\tilde{\nabla}W\,,\hspace*{10pt}
&\tilde{\nabla}_{(\mu}F_{\nu)}=-\delta_{\mu\nu}\nabla W\,,\nonumber\\
\nabla F_{\mu}=\pm\frac{1}{2}\tilde{\nabla}_{\mu}W\,,\hspace*{10pt}
\tilde{\nabla}F_{\mu}=\mp&\frac{1}{2}\nabla_{\mu}W\,,\hspace*{10pt}
\nabla_{\mu\nu}W=-\epsilon_{\mu\nu\rho\sigma}\nabla_{\rho}F_{\sigma}\,,\nonumber\\
\nabla_{[\mu}F_{\nu]}=\epsilon_{\mu\nu\rho\sigma}\tilde{\nabla}_{\rho}F_{\sigma}\,,\hspace*{10pt}
\tilde{\nabla}_{[\mu}F_{\nu]}=\epsilon_{\mu\nu\rho\sigma}\nabla_{\rho}F_{\sigma}&\,,\hspace*{10pt}
\nabla_{\mu\nu}F_{\rho}=-\delta_{\mu\nu\rho\sigma}\nabla F_{\sigma}+\epsilon_{\mu\nu\rho\sigma}\tilde{\nabla}F_{\sigma}\,,\nonumber\\
F_{\underline{\mu}}=-i\nabla F_{\mu}\,,\hspace*{10pt}
\tilde{F}_{\underline{\mu}}=i\tilde{\nabla}F_{\mu}\,,\hspace*{10pt}
&F_{\mu\underline{\nu}}=-i\nabla_{\nu}F_{\mu}\,,\hspace*{10pt}
\tilde{F}_{\mu\underline{\nu}}=i\tilde{\nabla}_{\nu}F_{\mu}\,,\nonumber\\
F_{\mu\nu\underline{\rho}}=-i(\delta_{\mu\nu\rho\sigma}\nabla F_{\sigma}+\epsilon_{\mu\nu\rho\sigma}\tilde{\nabla}F_{\sigma})\,,\hspace*{10pt}
&F_{\underline{\mu}\underline{\nu}}=\nabla_{[\mu}\nabla F_{\nu]}+i[F_{\mu},F_{\nu}]\,,\nonumber\\
G=\tilde{G}=G_{\mu\nu}=0\,,\hspace*{10pt}
G_{\mu}=2\tilde{\nabla}F_{\mu}\,,\hspace*{10pt}
&\tilde{G}_{\mu}=-2\nabla F_{\mu}\,,\hspace*{10pt}
G_{\underline{\mu}}=\frac{i}{2}(\nabla
G_{\mu}+\tilde{\nabla}\tilde{G}_{\mu})\,,\nonumber\\
ZF_{\mu}=\frac{1}{2}(\nabla G_{\mu}
-\tilde{\nabla}\tilde{G}_{\mu})\,,\hspace*{10pt}
&ZW=2i\nabla_{\underline{\mu}}
F_{\mu}+2\nabla\tilde{\nabla}W \nonumber
\end{align}\\

We define component fields of N=D=4 super multiplets as \\
\begin{align}
F_{\mu}|=\phi_{\mu}\,,\hspace*{10pt}
W|=A\,,\hspace*{10pt}
\nabla F_{\mu}|=\lambda_{\mu}\,,\hspace*{10pt}
&\tilde{\nabla}F_{\mu}|=\tilde{\lambda}_{\mu}\,,\hspace*{10pt}
\nabla_{\mu}F_{\nu}|=\delta_{\mu\nu}\rho+\rho_{\mu\nu}\,,\nonumber\\
\tilde{\nabla}_{\mu}F_{\nu}|=\delta_{\mu\nu}\tilde{\rho}+\frac{1}{2}\epsilon_{\mu\nu\rho\sigma}\rho_{\rho\sigma}\,,\hspace*{10pt}
\nabla\tilde{\nabla}W|=H\,,&\hspace*{10pt}
G_{\underline{\mu}}|=g_{\mu}\,,\hspace*{10pt}
ZF_{\mu}|=H_{\mu}\,,\hspace*{10pt}
ZW|=K.\notag
\end{align}
Twisted SUSY transformation of these component fields are given by\\
\begin{center}
\begin{tabular}{c|c|c}
&$s$&$\tilde{s}$ \\  \hline
 $\phi_{\rho}$&$\lambda_{\rho}$&$\tilde{\lambda}_{\rho}$ \\[2pt]
 $A_{\rho}$&$i\tilde{\lambda}_{\rho}$&$-i\lambda_{\rho}$ \\[2pt]
$A^{\prime}$&$\mp\tilde{\rho}$&$\pm\rho$ \\[2pt]
$\lambda_{\rho}$&$H^{\prime}_{\rho}$&$g^{\prime}_{\rho}$ \\[2pt]
$\tilde{\lambda}_{\rho}$&$-g^{\prime}_{\rho}$&$H^{\prime}_{\rho}$ \\[2pt]
 $\rho$&$-\frac{i}{2}D_{\mu}\phi_{\mu}$&$\pm K^{\prime}$ \\[2pt]
 $\tilde{\rho}$&$\mp K^{\prime}$&$-\frac{i}{2}D_{\mu}\phi_{\mu}$ \\[2pt]
 $\rho_{\rho\sigma}$&$-\frac{1}{2}\epsilon_{\rho\sigma\alpha\beta}F^{-}_{\alpha\beta}-\frac{i}{2}D_{[\rho}\phi_{\sigma]}$&$F^{-}_{\rho\sigma}-\frac{i}{2}\epsilon_{\rho\sigma\alpha\beta}D_{\alpha}\phi_{\beta}$ \\[2pt]
$K^{\prime}$&$\pm\frac{i}{2}(D_{\mu}\tilde{\lambda}_{\mu}+[\phi_{\mu},\lambda_{\mu}])$&$\mp\frac{i}{2}(D_{\mu}\lambda_{\mu}-[\phi_{\mu},\tilde{\lambda}_{\mu}])$ \\[2pt]
 $g^{\prime}_{\rho}$&$\pm\frac{i}{2}(D_{\rho}\tilde{\rho}-\frac{1}{2}\epsilon_{\rho\sigma\alpha\beta}D_{\sigma}\rho_{\alpha\beta}$&$\mp\frac{i}{2}(D_{\rho}\rho-D_{\sigma}\rho_{\rho\sigma}+[\phi_{\rho},\tilde{\rho}]$ \\
&$-[\phi_{\rho},\rho]-[\phi_{\sigma},\rho_{\rho\sigma}]\pm
 2[A^{\prime},\tilde{\lambda}_{\rho}])$&$+\frac{1}{2}\epsilon_{\rho\sigma\alpha\beta}[\phi_{\sigma},\rho_{\alpha\beta}]\pm
 2[A^{\prime},\lambda_{\rho}])$ \\[2pt]
 $H^{\prime}_{\rho}$&$\mp\frac{i}{2}(D_{\rho}\rho-D_{\sigma}\rho_{\rho\sigma}+[\phi_{\rho},\tilde{\rho}]$&$\mp\frac{i}{2}(D_{\rho}\tilde{\rho}-\frac{1}{2}\epsilon_{\rho\sigma\alpha\beta}D_{\sigma}\rho_{\alpha\beta}$ \\
&$+\frac{1}{2}\epsilon_{\rho\sigma\alpha\beta}[\phi_{\sigma},\rho_{\alpha\beta}]\pm
 2[A^{\prime},\lambda_{\rho}])$&$-[\phi_{\rho},\rho]-[\phi_{\sigma},\rho_{\rho\sigma}]\pm
 2[A^{\prime},\tilde{\lambda}_{\rho}])$
\end{tabular}\\
\end{center}

\begin{center}
\begin{tabular}{c|c|c}
&$s_{\mu}$&$\tilde{s}_{\mu}$  \\  \hline
 $\phi_{\rho}$&$\delta_{\mu\rho}\rho+\rho_{\mu\rho}$&$\delta_{\mu\rho}\tilde{\rho}+\frac{1}{2}\epsilon_{\mu\rho\alpha\beta}\rho_{\alpha\beta}$ \\[2pt]
 $A_{\rho}$&$-i\delta_{\mu\rho}\tilde{\rho}+\frac{i}{2}\epsilon_{\mu\rho\alpha\beta}\rho_{\alpha\beta}$&$i\delta_{\mu\rho}\rho-i\rho_{\mu\rho}$ \\[2pt]
$A^{\prime}$&$-\tilde{\lambda}_{\mu}$&$\lambda_[\mu]$ \\[2pt]
$\lambda_{\rho}$&$\frac{1}{2}\epsilon_{\mu\rho\alpha\beta}F^{-}_{\alpha\beta}+\frac{i}{2}\delta_{\mu\rho}D_{\nu}\phi_{\nu}-\frac{i}{2}D_{(\mu}\phi_{\rho)}$&$F^{+}_{\mu\rho}\pm\delta_{\mu\rho}K^{\prime}+\frac{i}{2}\epsilon_{\mu\rho\alpha\beta}D_{\alpha}\phi_{\beta}$ \\[2pt]
$\tilde{\lambda}_{\rho}$&$-F^{+}_{\mu\rho}\pm\delta_{\mu\rho}K^{\prime}+\frac{i}{2}\epsilon_{\mu\rho\alpha\beta}D_{\alpha}\phi_{\beta}$&$-\frac{1}{2}\epsilon_{\mu\rho\alpha\beta}F^{-}_{\alpha\beta}+\delta_{\mu\rho}\frac{i}{2}D_{\nu}\phi_{\nu}-\frac{i}{2}D_{(\mu}\phi_{\rho)}$ \\[2pt]
 $\rho$&$-\frac{i}{2}D_{\mu}\phi_{\mu}$&$\mp\frac{i}{2}g_{\mu}$ \\[2pt]
 $\tilde{\rho}$&$\pm g_{\mu}$&$\frac{1}{2}H_{\mu}$ \\[2pt]
 $\rho_{\rho\sigma}$&$\pm\frac{1}{2}\delta_{\mu\nu\rho\sigma}H_{\nu}\mp\frac{i}{2}\epsilon_{\mu\nu\rho\sigma}g_{\nu}$&$\pm\frac{1}{2}\epsilon_{\mu\nu\rho\sigma}H_{\nu}\pm\frac{i}{2}\delta_{\mu\nu\rho\sigma}g_{\nu}$ \\[2pt]
$K^{\prime}$&$\pm\frac{i}{2}(D_{\mu}\tilde{\rho}-\frac{1}{2}\epsilon_{\mu\nu\rho\sigma}D_{\nu}\rho_{\rho\sigma}-[\phi_{\mu},\rho]-[\phi_{\nu},\rho_{\mu\nu}])$&$\mp\frac{i}{2}(D_{\mu}\rho-D_{\nu}\rho_{\mu\nu}+[\phi_{\mu},\tilde{\rho}]+\frac{1}{2}\epsilon_{\mu\nu\rho\sigma}[\phi_{\nu},\rho_{\rho\sigma}])$ \\[2pt]
 $g^{\prime}_{\rho}$&$-\frac{i}{2}(\delta_{\mu\rho}(D_{\nu}\tilde{\lambda}_{\nu}+[\phi_{\nu},\lambda_{\nu}])-D_{(\mu}\tilde{\lambda}_{\rho)}$&$-\frac{i}{2}(-\delta_{\mu\rho}(D_{\nu}\lambda_{\nu}-[\phi_{\nu},\tilde{\lambda}_{\nu}])+D_{(\mu}\lambda_{\rho)}$ \\
&$+[\phi_{[\mu},\lambda_{\rho]}]-\epsilon_{\mu\rho\alpha\beta}(D_{\alpha}\lambda_{\beta}+[\phi_{\alpha},\tilde{\lambda}_{\beta}]))$&$+[\phi_{[\mu},\tilde{\lambda}_{\rho]}]+\epsilon_{\mu\rho\alpha\beta}(D_{\alpha}\tilde{\lambda}_{\beta}-[\phi_{\alpha},\lambda_{\beta}]))$ \\[2pt]
 $H^{\prime}_{\rho}$&$-\frac{i}{2}(\delta_{\mu\rho}(D_{\nu}\lambda_{\nu}-[\phi_{\nu},\tilde{\lambda}_{\nu}])+D_{[\mu}\lambda_{\rho]}$&$-\frac{i}{2}(\delta_{\mu\rho}(D_{\nu}\tilde{\lambda}_{\nu}+[\phi_{\nu},\lambda_{\nu}])+D_{[\mu}\tilde{\lambda}_{\rho]}$ \\
&$+[\phi_{(\mu},\tilde{\lambda}_{\rho)}]+\epsilon_{\mu\rho\alpha\beta}(D_{\alpha}\tilde{\lambda}_{\beta}-[\phi_{\alpha},\lambda_{\beta}]))$&$-[\phi_{(\mu},\lambda_{\rho)}]+\epsilon_{\mu\rho\alpha\beta}(D_{\alpha}\lambda_{\beta}+[\phi_{\alpha},\tilde{\lambda}_{\beta}]))$
\end{tabular}
\end{center}

\begin{center}
\begin{tabular}{c|c|c}
&$s_{\mu\nu}$&$Z$ \\[2pt] \hline
$\phi_{\rho}$&$-\delta_{\mu\nu\rho\sigma}\lambda_{\sigma}+\epsilon_{\mu\nu\rho\sigma}\tilde{\lambda}_{\sigma}$&$H_{\rho}$ \\[2pt]
$A_{\rho}$&$i\delta_{\mu\nu\rho\sigma}\tilde{\lambda}_{\sigma}+i\epsilon_{\mu\nu\rho\sigma}\lambda_{\sigma}$&$g_{\rho}$ \\[2pt]
$A^{\prime}$&$\mp\frac{1}{2}\epsilon_{\mu\nu\rho\sigma}\rho_{\rho\sigma}$&$2K^{\prime}$ \\[2pt]
$\lambda_{\rho}$&$\delta_{\mu\nu\rho\sigma}H^{\prime}_{\sigma}+\epsilon_{\mu\nu\rho\sigma}g^{\prime}_{\sigma}$&$\mp i(D_{\rho}\rho-D_{\sigma}\rho_{\rho\sigma}+[\phi_{\rho},\tilde{\rho}]$ \\[2pt]
&&$+\frac{1}{2}\epsilon_{\rho\sigma\alpha\beta}[\phi_{\sigma},\rho_{\alpha\beta}])$ \\ 
$\tilde{\lambda}_{\rho}$&$\delta_{\mu\nu\rho\sigma}g^{\prime}_{\sigma}-\epsilon_{\mu\nu\rho\sigma}H^{\prime}_{\sigma}$&$\mp i(D_{\rho}\tilde{\rho}-\frac{1}{2}\epsilon_{\rho\sigma\alpha\beta}D_{\sigma}\rho_{\alpha\beta}-[\phi_{\rho},\rho]$ \\[2pt]
&&$-[\phi_{\sigma},\rho_{\rho\sigma}])$\\
$\rho$&$\frac{1}{2}\epsilon_{\mu\nu\rho\sigma}F^{-}_{\rho\sigma}-\frac{i}{2}D_{[\mu}\phi_{\nu]}$&$-i(D_{\rho}\lambda_{\rho}-[\phi_{\rho},\tilde{\lambda}_{\rho}]-i[A,\rho])$ \\[2pt]
$\tilde{\rho}$&$F^{-}_{\mu\nu}+\frac{i}{2}\epsilon_{\mu\nu\rho\sigma}D_{\rho}\phi_{\sigma}$&$-i(D_{\rho}\tilde{\lambda}_{\rho}+[\phi_{\rho},\lambda_{\rho}]-[A,\tilde{\rho}])$ \\[2pt]
$\rho_{\rho\sigma}$&$\epsilon_{\mu\nu\alpha\gamma}\delta_{\rho\sigma\beta\gamma}F^{+}_{\alpha\beta}+\frac{i}{2}\delta_{\mu\nu\alpha\gamma}\delta_{\rho\sigma\beta\gamma}D_{(\alpha}\phi_{\beta)}$&$-i(D_{[\mu}\lambda_{\nu]}+[\phi_{[\mu},\tilde{\lambda}_{\nu]}]$ \\
&$-\frac{i}{2}\delta_{\mu\nu\rho\sigma}D_{\alpha}\phi_{\alpha}\mp\epsilon_{\mu\nu\rho\sigma}K^{\prime}$&$+\epsilon_{\mu\nu\rho\sigma}(D_{\rho}\tilde{\lambda}_{\sigma}-[\phi_{\rho},\lambda_{\sigma}])-[A,\rho_{\mu\nu}])$\\[2pt]
$K^{\prime}$&$\pm\frac{i}{2}(D_{[\mu}\tilde{\lambda}_{\nu]}-[\phi_{[\mu},\lambda_{\nu]}]+\epsilon_{\mu\nu\rho\sigma}(D_{\rho}\lambda_{\sigma}$&$\pm i(\{\lambda,\lambda\}+\{\tilde{\lambda},\tilde{\lambda}\}-D_{\rho}g^{\prime}_{\rho}+[\phi_{\rho},H^{\prime}_{\rho}]$ \\[2pt]
&&$\pm[A,K^{\prime}])$\\
$g^{\prime}_{\rho}$&$\pm\frac{i}{2}\delta_{\mu\nu\rho\sigma}(D_{\sigma}\tilde{\rho}-\frac{1}{2}\epsilon_{\sigma\gamma\alpha\beta}D_{\gamma}\rho_{\alpha\beta}-[\phi_{\sigma},\rho]$&$\mp i(-\frac{1}{2}D_{\sigma}F_{\rho\sigma}\pm D_{\rho}K^{\prime}-\frac{i}{2}[\phi_{\sigma},D_{\rho}\phi_{\sigma}]$
 \\[2pt]
&$+[\phi_{rho},\tilde{\lambda}_{\sigma}]))-[\phi_{\alpha},\rho_{\sigma\alpha}]\pm[A,\tilde{\lambda}_{\sigma}])\pm\frac{i}{2}\epsilon_{\mu\nu\rho\sigma}(D_{\sigma}\rho$
&$+\{\rho,\lambda_{\rho}\}+\{\tilde{\rho},\tilde{\lambda}_{\rho}\}+\{\lambda_{\sigma},\rho_{\rho\sigma}\}$ \\
&$-D_{\alpha}\rho_{\sigma\alpha}+[\phi_{\sigma},\tilde{\rho}]+\frac{1}{2}\epsilon_{\sigma\gamma\alpha\beta}[\phi_{\gamma},\rho_{\alpha\beta}]\pm[A,\lambda_{\sigma}])$&$+\frac{1}{2}\epsilon_{\rho\sigma\alpha\beta}\{\tilde{\lambda}_{\sigma},\rho_{\alpha\beta}\})$\\
$H^{\prime}_{\rho}$&$\pm\frac{i}{2}\delta_{\mu\nu\rho\sigma}(D_{\sigma}\rho-D_{\alpha}\rho_{\sigma\alpha}+[\phi_{\sigma},\tilde{\rho}]$&$\mp i(-\frac{i}{2}D_{\sigma}D_{\sigma}\phi_{\rho}+\frac{i}{2}[\phi_{\sigma},[\phi_{\rho},\phi_{\sigma}]]$\\ 
&$+\frac{1}{2}\epsilon_{\sigma\gamma\alpha\beta}[\phi_{\gamma},\rho_{\alpha\beta}]\pm[A,\lambda_{\sigma}])\mp\frac{i}{2}\epsilon_{\mu\nu\rho\sigma}(D_{\sigma}\tilde{\rho}$&$\mp[\phi_{\rho},K^{\prime}]+\{\rho,\tilde{\lambda}_{\rho}\}-\{\tilde{\rho},\lambda_{\rho}\}$ \\
&$-\frac{1}{2}\epsilon_{\sigma\gamma\alpha\beta}D_{\gamma}\rho_{\alpha\beta}-[\phi_{\sigma},\rho]-[\phi_{\alpha},\rho_{\sigma\alpha}]\pm[A,\tilde{\lambda}_{\sigma}])$&$-\{\tilde{\lambda}_{\sigma},\rho_{\rho\sigma}\}+\frac{1}{2}\epsilon_{\rho\sigma\alpha\beta}\{\lambda_{\sigma},\rho_{\alpha\beta}\})$
\end{tabular}
\end{center}

For a closure of SUSY algebra for this ansatz we again need the following 
constraint:
\begin{align}
&iD_{\mu}g_{\mu}-[\phi_{\mu},H_{\mu}]-\frac{i}{2}D_{\mu}D_{\mu}A-
\frac{i}{2}[\phi_{\mu},[\phi_{\mu},A]]\mp\frac{1}{4}[A,K]\nonumber \\
&\mp\{\rho,\rho\}\mp\{\tilde{\rho},\tilde{\rho}\}\mp\frac{1}{2}\{\rho_{\mu\nu},\rho_{\mu\nu}\}-\{\lambda_{\mu},\lambda_{\mu}\}-\{\tilde{\lambda}_{\mu},\tilde{\lambda}_{\mu}\}=0.
\end{align}
 For the case of one central charge we have tried all possible 
super connection ansatz. We have found out two possible 
consistent Ansatz (B) and (C) but for both cases we need a constraint 
equation for the off-shell closure of N=D=4 twisted algebra 
with a central charge. 
Off-shell N=D=4 twisted SUSY invariant action for this Ansatz (C) can be 
given by 
\begin{equation}
\begin{split}
S=&\displaystyle\int
 d^{4}x\text{Tr}\Bigl(\dfrac{1}{2}D_{\mu}\phi_{\nu}D_{\mu}\phi_{\nu}
-\frac{1}{4}F^{2}_{\mu\nu}\pm\frac{1}{2}(g_{\mu}-D_{\mu}A)^{2}\mp\frac{1}{2}(H_{\mu}-i[A,\phi_{\mu}])^{2}-\frac{1}{8}K^{2} \\
&-2i\rho(D_{\mu}\lambda_{\mu}-[\phi_{\mu},\tilde{\lambda}_{\mu}])
-2i\tilde{\rho}(D_{\mu}\tilde{\lambda}_{\mu}+[\phi_{\mu},\lambda_{\mu}])-2i\rho_{\mu\nu}(D_{\mu}\lambda_{\nu}+[\phi_{\mu},\tilde{\lambda}_{\nu}]) \\
&-i\epsilon_{\mu\nu\alpha\beta}\rho_{\mu\nu}(D_{\alpha}\tilde{\lambda}_{\beta}-[\phi_{\alpha},\lambda_{\beta}])
\pm iA(\{\lambda_{\mu},\lambda_{\mu}\}+\{\tilde{\lambda}_{\mu},\tilde{\lambda}_{\mu}\})
+\frac{1}{4}[\phi_{\mu},\phi_{\nu}]^{2}\Bigl). 
\end{split}
\end{equation}

\section{Equivalence of the Ansatz (B) and Ansatz (C)}

Consider the following general Ansatz of N=D=4:\\
\begin{center}
\begin{tabular}{c|c|c|c|c}
&$\nabla$&$\tilde{\nabla}$&$\nabla_{\rho}$&$\tilde{\nabla}_{\rho}$ \\ \hline
$\nabla$&$X_{0}+X^{\prime}_{5}$&$X_{5}+X^{\prime}_{0}$&$-i(\nabla_{\underline{\rho}}+iX_{\rho})$&$-X^{\prime}_{\rho}$ \\
$\tilde{\nabla}$&&$X_{0}+X^{\prime}_{5}$&$X^{\prime}_{\rho}$&$-i(\nabla_{\underline{\rho}}-iX_{\rho})$ \\
$\nabla_{\mu}$&&&$\delta_{\mu\rho}(X_{0}-X^{\prime}_{5})$&$\delta_{\mu\rho}(X_{5}-X^{\prime}_{0})$ \\
$\tilde{\nabla}_{\mu}$&&&&$\delta_{\mu\rho}(X_{0}-X^{\prime}_{5})$ \\
$\nabla_{\mu\nu}$&&&& \\
$\nabla_{\underline{\mu}}$&&&& \\
\end{tabular}
\\
\begin{tabular}{c|c|c|c}
&$\nabla_{\rho\sigma}$&$\nabla_{\underline{\rho}}$&$\nabla_{z}$ \\ \hline
$\nabla$&$0$&$-iF_{\underline{\rho}}$&$iG$ \\
$\tilde{\nabla}$&$0$&$-i\tilde{F_{\underline{\rho}}}$&$i\tilde{G}$ \\
$\nabla_{\mu}$&$i\delta_{\mu\nu\rho\sigma}(\nabla_{\underline{\nu}}-iX_{\nu})-\epsilon_{\mu\nu\rho\sigma}X^{\prime}_{\nu}$&$-iF_{\mu\underline{\rho}}$&$iG_{\mu}$ \\
$\tilde{\nabla}_{\mu}$&$-i\epsilon_{\mu\nu\rho\sigma}(\nabla_{\underline{\nu}}+iX_{\nu})-\delta_{\mu\nu\rho\sigma}X^{\prime}_{\nu}$&$-i\tilde{F}_{\mu\underline{\rho}}$&$i\tilde{G}_{\mu}$ \\
$\nabla_{\mu\nu}$&$\delta_{\mu\nu\rho\sigma}(X_{0}+X^{\prime}_{5})-\epsilon_{\mu\nu\rho\sigma}(X_{5}+X^{\prime}_{0})$&$-iF_{\mu\nu\underline{\rho}}$&$iG_{\mu\nu}$ \\
$\nabla_{\underline{\mu}}$&&$-iF_{\underline{\mu}\underline{\rho}}$&$iG_{\underline{\mu}}$ \\
$\nabla_{z}$&&&$0$
\end{tabular}\\
Ansatz (D)
\end{center}
Define the following new connections and curvatures:
\begin{equation}
\begin{split}
\nabla^{\text{new}}&=\frac{1}{\sqrt{2}}(-i\nabla+\tilde{\nabla})\,,
\tilde{\nabla}^{\text{new}}=\frac{1}{\sqrt{2}}(\nabla-i\tilde{\nabla})\,,
\\
 \nabla_{\mu}^{\text{new}}&=\frac{1}{\sqrt{2}}(i\nabla_{\mu}+\tilde{\nabla}_{\mu})\,,
\tilde{\nabla}_{\mu}^{\text{new}}=\frac{1}{\sqrt{2}}(\nabla_{\mu}+i\tilde{\nabla}_{\mu})\,,
\\
 &\nabla_{\mu\nu}^{\text{new}}=\frac{1}{\sqrt{2}}(-i\nabla_{\mu\nu}-\frac{1}{2}\epsilon_{\mu\nu\rho\sigma}\nabla_{\rho\sigma})\,,
\end{split}
\end{equation}
\begin{equation}
\begin{split}
&X_{0}^{\hbox{new}}= -iX^{\prime}_{0}\,,\hspace*{10pt} 
X_{\mu}^{\hbox{new}}=iX_{\mu}^{\prime}\,,\hspace*{10pt} 
X_{5}^{\hbox{new}}= -iX^{\prime}_{5}\\
&X_{0}^{\prime\,\hbox{new}}= -iX_{0}\,,\hspace*{10pt}
X_{\mu}^{\prime\,\hbox{new}}=iX_{\mu}\,,\hspace*{10pt}
X_{5}^{\prime\,\hbox{new}}= -iX_{5}
\end{split}
\end{equation}

It turns out that this Ansatz (D) and the ansatz given by the above new 
system has exactly the same form. Surprisingly the Ansatz (B) and 
Ansatz (C) have exactly the same relations as this new and old system. 
In other words these ansatz are essentially the same and thus both 
of cases naturally need the constraint relation for the N=4 SUSY closure.   

\section{Conclusion and Discussions}

In 2-dimension we found two types of super connection ansatz which 
realize off-shell closure of N=2 twisted SUSY including central charge 
with and without constraint. 
Off-shell twisted SUSY invariant actions are found 
for each ansatz. On the other hand in 4-dimension we examined 
two possible ansatz of N=4 twisted SUSY algebra with a central 
charge and we found that both of ansatz having similarity with 
the 2-dimensional ansatz need a constraint equation for the 
off-shell closure of N=4 twisted SUSY algebra. In fact we found 
that these two ansatz are essentially equivalent to each other. 

We have thus investigated a possibility of off-shell twisted N=D=4 
invariant super Yang-Mills formulation without a constraint by 
super connection formalism. As far as N=4 twisted SUSY algebra with 
one super charge is concerned a constraint is inevitable for the 
off-shell closure of the algebra. 
We consider that this may be related to the fact that 
10-dimensional N=1 super Yang-Mills theory can be formulated only 
at the on-shell level. N=D=4 super Yang-Mills can be dimensionally 
reduced from the 10-dimensional N=1 formulation and the on-shell 
nature could be kept invariant in the dimensionally reduced 
formulation and thus may lead to off-shell closure but with 
a constraint in 4-dimensions.

\vspace{1cm}
{\bf{\Large Acknowledgements}}

I would like to thank my collaborators, K. Asaka, A. D'Adda, I. Kanamori, 
J. Kato, A. Miyake, K. Nagata, J. Saito, T. Tsukioka and Y. Uchida for 
useful discussions and fruitful collaborations. 
Thanks are also due to I. L. Buchbinder and V. Epp for useful comments 
and kind hospitality in Tomsk. 
This work was supported in part by Japanese Ministry of Education,
Science, Sports and Culture under the grant number 22540261.

\end{document}